\documentclass{kapproc}
\usepackage{procps}
\usepackage[dvips]{graphicx}
\upperandlowercase
\begin{document}

\articletitle[Generation of the Hubble Sequence]
{On the Generation of\\
the Hubble Sequence through an \\
Internal Secular Dynamical Process}
\author{Xiaolei Zhang\altaffilmark{1}}
 
\affil{\altaffilmark{1}US Naval Rsearch Laboratory,
4555 Overlook Ave. SW, Washington, DC 20375, USA}

\begin{abstract}
The secular evolution process, which slowly transforms
the morphology of a galaxy over its lifetime, could naturally 
account for observed properties of the great majority
of physical galaxies if both stellar and gaseous accretion processes
are taken into account.  As an emerging paradigm for galaxy evolution, 
its dynamical foundation had been established in the past few years, 
and its observational consequences are yet to be fully explored.
The secular evolution picture provides a coherent framework for 
understanding the extraordinary regularity and the systematic variation 
of galaxy properties along the Hubble sequence.
\end{abstract}


\section{Introduction}

The dominant view over the past few decades has been that the 
structural properties of galaxies remain largely unchanged
unless galaxies were perturbed by violent events such as mergers
(Toomre \& Toomre 1972).  This view is particularly
favored by the currently popular hierarchical clustering/cold dark
matter (CDM) paradigm of structure formation and evolution 
(Peebles 1993 and the references therein).  In the early 1980s, 
photometric and kinematic evidence in the buldges of late-type 
galaxies, and hints from N-body simulations of barred 
galaxies which incorporated a dissipative gas component, prompted
several investigators to speculate that a fraction of late-type
bulges might be formed from gas accretion under the bar potential
(Kormendy 1982; Combes \& Sander 1981).  These early observations
and speculations have since been further developed into one version 
of the secular evolution scenario, which emphasizes the role of 
dissipative gas accretion in the formation of the so-called 
``pseudo bulges'' in late type galaxies (Kormendy \& Kennicutt 2004 
and the references therein).

Even though the role of mergers during the early phases of galaxy assembly is
still to be assessed, growing evidence has shown that at least since z $\sim$ 1
the rate of merger appears to have been significantly reduced
(Conselice et al. 2003), and during subsequent time the significance of 
merger in transforming galaxy morphology is likely to be overshadowed 
by the slower secular evolution process (Kormendy \& Kennicutt 2004). 

However, secular evolution involving gas alone under barred potential
leads to some apparent paradoxes.  First of all, as emphasized
by Andredakis and coworkers (Andredakis, Peletier, \& Balcells 1995),
the continuity of the galaxy properties across the entire Hubble sequence,
highlighted for example by the continued variation of Sersic index n
in fitting the bulge surface density profile, indicates that there is not 
an apparent break in the formation mechanism between the late-type disk 
galaxies and mid-to-early-type disk galaxies. However, due to the paucity of
gas compared to stars in most galaxies, manifesting as a gaseous-to-stellar 
mass ratio of 1/10 or less (which is true even for high-redshift
late-type disk galaxies, presumably a result of rapid star formation 
following the accretion shocks of dissipative disk formation, D. Sanders
2002 private communication; F. Combes 2004 private communication),
there simply is not enough of a reservoir of gas for use to build up 
the bulges of intermediate Hubble type galaxies such as our own, which
has a bulge mass comparable to the disk mass, not to say for galaxies 
of even earlier Hubble types. Secondly, bulges of intermediate to 
early type galaxies, including our own, consists mostly of stars 
of very old age (Jablonka, Gorgas \& Goudfrooij 2002), and could not 
have been built up by the secular accretion of gas over a Hubble time
which subsequently formed stars.  Even though Kormendy and Kennicutt (2004)
argued that some of these bulges may be termed ``classical'', i.e., they
could have formed out of the dissipative collapse at an early stage of
galaxy formation, they nonetheless pointed out that a good fraction of these
apparently old bulges have stellar kinematics which were rotation-dominated,
and are related to the kinematics of their disks, hinting at a
secular evolution origin for their formation.

These and other apparent paradoxes, such as the existence in some
galaxies of dense core in the central region which appears to be formed by 
radial gas accretion and yet is kinematically decoupled from the bulge stars, 
can be naturally resolved if we allow also the possibility of stellar accretion.
Such a possibility was not explored in the past decades mainly due to a 
theoretical barrier, i.e., the well-known result that the stellar motion in a 
galaxy containing quasi-stationary non-axisymmetric patterns conserves 
the Jacobi integral in the rotating frame of the pattern (Binney 
\& Tremaine 1987), and the orbital motion of stars under such potentials
generally will not exhibit secular delay or increase, since there is no wave
and disk star interaction except at the wave-particle resonances
which are usually localized for quasi-steady wave patterns
(Lynden-Bell \& Kalnajs 1972).

\section{Physical Mechanisms for Producing the
Secular Morphological Evolution of Galaxies}

It was first demonstrated in Zhang (1996, 1998, 1999) that secular orbital
changes of stars across the entire galaxy disk are in fact possible due to the 
collective instabilities induced by the unstable density wave modes
such spirals, bars, as well as by the skewed 3D density distributions
reflected in the twisted isophotes of many high-redshift galaxies.  
These skewed global patterns were shown to lead to a secular energy 
and angular momentum exchange process between the disk matter and 
the wave pattern, mediated by a local gravitational instability,
or a collisionless shock, at the potential minimum of the pattern 
(Zhang 1996).  The integral manifestation of this process 
is an azimuthal phase shift between the potential and density spirals,
which results in a secular torque action between the wave pattern and the
underlying disk matter.  As a result of the torquing of the wave on the 
disk matter, the matter inside the corotation radius loses angular momentum 
to the wave secularly and sinks inward.  The wave carries the angular momentum 
it receives from the inner disk matter to the outer disk and deposits 
it there, causing the matter in the outer disk to drift further out.  

Closed form evolution rates can be derived for this process.
In the quasi steady state, the rate of angular momentum exchange
between a skewed density wave pattern (spiral, bar, etc.)
and the basic state of the disk,
per unit area, is given by (Zhang 1996, 1998)
\begin{equation}
{\overline{ {{dL} \over {dt}}}} (r)
=  - { {1} \over {2 \pi} }
\int_0^{2  \pi}
\Sigma_1(r,\phi)
{ {\partial {{{\cal V}_1}(r,\phi)}} \over {\partial \phi}}
d \phi
,
\end{equation}
which, for two sinusoidal waveforms, is given by
\begin{equation}
{\overline{ {{dL} \over {dt}}}} (r)
=
(m/2) A_{\Sigma} A_{\cal{V}} sin (m \phi_0)
,
\label{eq:A}
\end{equation}
where $A_{\Sigma}$ and $A_{\cal{V}}$ are the amplitudes
of the density and potential waves, respectively,
and $\phi_0$ is the phase shift between these two
waveforms.

The orbital decay rate it induced on a disk star 
(or a gas clump) can be
derived from equation (\ref{eq:A}) as
\begin{equation}
{{dr} \over {dt}}
= - { 1 \over 2} F^2 v_0 \tan (i) \sin(m \phi_0)
,
\label{eq:decay}
\end{equation}
where $F$ is the fractional wave amplitude (which
is the geometric average of the fractional density and potential
wave amplitude), $v_0$ is the circular
velocity of the star, $i$ is the pitch angle of the spiral.
$m$ is the number of spiral arms, respectively.  

From the above expressions we can easily see why the skewness
of the pattern is needed: It is the skewness of the mass distribution
which leads to the potential/density phaseshift through the
Poisson equation (Zhang 1996).  Without phaseshift,
there will be no secular angular momentum exchange between
the wave and the disk matter at the quasi-steady state, and thus
no secular mass redistribution.  A perfect oval or triaxial
mass distribution without any skewness will not be the direct
driver for secular evolution.

For our own Galaxy, 
if we assume that over the past Hubble time the Galactic spiral
pattern has an average 20\% amplitude and 20 degree pitch angle,
the orbital delay rate can be calculated using equation (\ref{eq:decay})
to be about 2 kpc of orbital decay per Hubble time,
Therefore, a star in the Sun's orbit will not make it all the way
in to the inner Galaxy in a Hubble time.
However, the corresponding mass accretion rate across any
Galactic radius inside corotation is about
$6 \times 10^9 M_{\odot}$ per Hubble time.
A substantial fraction of the Galactic Bulge can thus be built up
in a Hubble time.

Another important consequence of spiral-induced wave-basic state
interaction is the secular heating of the disk stars.
Since a spiral density wave can only gain energy and angular momentum
in proportion to the pattern speed $\Omega_p$,
and a disk star loses its orbital energy and angular momentum
in proportion to its circular speed $\Omega$,
an average star cannot lose the orbital energy entirely to the wave,
the excess energy serves to heat the star.
The resulting orbital velocity diffusion rate is given by
\begin{equation}
D = (\Omega - \Omega_p) F^2 v_c^2 \tan (i) \sin(m \phi_0 )
\label{eq:D}
.
\end{equation}
Using parameters appropriate for the Galaxy, this can
quantitatively explain the observed age-velocity dispersion relation
of the solar neighborhood stars (Zhang 1999; Wielen 1977).
This secular heating process, since it originates
from a local gravitational instability at the arms, serves to
increase all three space velocities of stars.  The vertical component 
of the velocity dispersion allows the stars to gradually drift out of the 
galactic plane as they spiral inward to become bulge stars.
The corresponding energy injection into the interstellar medium, if used
as the top-level energy input for turbulence cascade,
can quantitatively account for the size-linewidth relation of the
Galactic molecular clouds (Zhang et al. 2001; Zhang 2002).

The secular morphological evolution process leads to the Hubble type
of an average galaxy to evolve from the late to the early (Zhang 1999).
The secular evolution speed is expected to be faster for cluster
galaxies than for isolated field galaxies, which were
observed to evolve slowly, because the rate of 
secular evolution is proportional to the wave amplitude squared 
and pattern pitch angle squared 
(see, e.g., equations \ref{eq:decay}, \ref{eq:D}), 
and cluster galaxies are found 
to have large amplitude and open spiral patterns excited through
the tidal interactions with neighboring galaxies and with the cluster
potential as a whole.  This naturally accounts for
the morphological Butcher-Oemler effect.

The secular evolution of galaxies through the internal process
mediated by skewed non-axisymmetric patterns is a rigorous consequence 
of the Newtonian dynamics and the assumption of global self-consistency.
The analytical equations for the evolution rates (equations
\ref{eq:decay} and \ref{eq:D}) are 
quantitatively confirmed in N-body simulations (Zhang 1998).  
Detailed comparison of the predictions of the secular evolution
theory with the observed galaxy properties and with other
theories of galaxy evolution can be found in (Zhang 2003).  
In what follows, we will highlight the
role of secular evolution in generating the structural properties
and scaling relations of galaxies.  

\section{Generation of the Structural Properties of Galaxies 
Along the Hubble Sequence through Secular Evolution}

The properties of disk galaxies vary systematically along the
Hubble sequence from the late to the early Hubble types.  
An important descriptor of the structural characteristics of
disk galaxies is the rotation curve.  After assembling hundreds 
of observed rotation curves for nearby disk galaxies,
Persic, Salucci, \& Stel (1996) found that these rotation curves
fall onto a two-dimensional surface in the three-parameter space
of normalized galactic radius, velocity, and the absolute magnitude.
Since for the nearby galaxies the variation
of the absolute magnitude corresponds to the variation
of Hubble type (i.e., bigger and brighter galaxies generally
have earlier Hubble types), this means that the rotation curve
shapes also vary systematically for galaxies of varying Hubble
types.  In Figure 1, we plot three typical rotation curves generated
using the data from Persic et al. (1996).

Using dynamical relations for equilibrium axisymmetric/spherically
symmetric mass distributions, i.e.
$
V(R)^2 = {{G M_{dyn} (R)} \over R}
,
$
as well as 
$
M_{dyn}(R) = \int_0^R \Sigma_{proj} (r) 2 \pi r dr
=\int_0^{R} \rho(r) 4 \pi r^2 dr 
,
$
where $M_{dyn} (R)$ is the dynamical or total mass within a galactic
radius R (including the contributions from both the luminous and
the dark components), and $\Sigma_{proj} (R)$ is the total projected 
surface density (disk plus spheroidal) at radius R, we can infer 
the underlying mass and surface density distributions corresponding
to the different regimes for each type of rotation curves (c.f. Figure 1). 

\begin{figure}[ht]
\vskip.2in
\centerline{\includegraphics[width=2in]{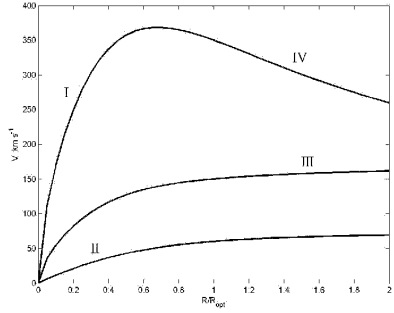}}
\caption{Typical rotation curves for disk galaxies of varying absolute
magnitudes, as well as four characteristic regimes which are
analyzed in the main text. Top: M=18; middle: M=21; bottom: M=24, where
M is the absolute magnitude of the representative galaxy (data from
Persic et al [1996]).}
\end{figure}

(I) Steeply rising solid body rotation curve, as observed
for the inner regions of early type galaxies.

This regime is characterized by solid body rotation ($\Omega
= \Omega_0$ = constant, or a nearly linear rise of rotation
velocity $V(R)=\Omega_0 R$).  
Using $V(R)^2 = \Omega_0^2 R^2 = GM_{dyn}/R$, we obtain
$
M_{dyn}= {{\Omega_0 R^3} \over {G}},
$
which implies a constant volume density
$
\rho = \rho_0 = 
{{\Omega_0 } \over {G}} {{3} \over {4 \pi}}.
$
Therefore the projected surface density can be obtained
by equating
$
M_{dyn} (R) = \int_0^R \Sigma_{proj} (r) 2 \pi r dr
$
and
$
M_{dyn} (R) = {{3} \over {4}} \pi R^3 \rho_0
,
$
which gives
$
\Sigma_{proj} (R) = 2 \rho_0 R
,
$
i.e. the solid-body regime corresponds to a constant column
density $\rho_0$ and linearly rising
projected surface density $\Sigma_{proj} (R) \propto R$.

(II) Gently rising rotation curve, as representative of the very late type 
disks, including most of the dwarf galaxies and
low-surface-brightness (LSB) disk galaxies.

This regime can be shown to correspond to approximately constant
projected surface density, as found to be the case for many LSB disks.
Assume $\Sigma_{proj} (R) = \Sigma_0$ = constant, we obtain
$
M_{dyn} (R) = \int_0^R \Sigma_0 2 \pi r dr = \Sigma_0 \pi R^2,
$
which we equate to $V(R)^2 R/G$ to obtain
$
V(R)^2 = \sqrt{ \Sigma_0 \pi R G}
,
$
which indeed corresponds to a gently rising rotation curve.

Instead of constant projected surface density, many LSBs are found
to have instead constant volume density (de Blok 2003).
These two views (constant projected surface density, or constant
mass volume density) can be consistent if the scale height
in the central regions of these LSBs are nearly constant, as
for a pancake-type proto galaxy cloud collapse.

(III) Flat rotation curve, as is representative of the outer part
of most Sb/Sc galaxies.

In this regime, the projected surface
density profile should be close to $1/R$, as shown below.
Assume 
$
\Sigma_{proj}(R) = {{\Sigma_0 R_0} \over R},
$
where $\Sigma_0 = \Sigma_{proj} (R=R_0)$, we have
$
M_{dyn} (R) = \int_0^R \Sigma_{proj} (r) 2 \pi r dr
$
$
= \Sigma_0 R_0 2 \pi R.
$
Using again $V(R)^2 = GM_{dyn}/R$, we obtain
$
V(R)^2 = G \Sigma_0 R_0 2 \pi R/R,
$
or
$
V(R) = \sqrt{ G \Sigma_0 R_0 2 \pi}$ = constant.
Therefore the flat rotation curve regime corresponds to 1/R 
projected surface density distribution.

(IV) The falling rotation curve regime, as representative of
the outer part of the early type galaxies.
 
In this regime the projected surface density $\Sigma_{proj} (R)$ 
falls off faster than 1/R, and can be up to $\Sigma_{proj} =0$ (Keplerian).
For the extreme case of Keplerian rotation curve due to
a concentrated central mass distribution, $\Sigma_{proj}(R_{outer}) \approx 0$,
and 
$
V(R)^2 = {{G M_{dyn}} \over R},
$
or
$
V(R)= \sqrt{ { G M_{dyn}} \over R},
$
which is indeed falling. 

In the secular evolution scenario a typical galaxy starts its
life with properties similar to that of an LSB galaxy.
The high-z correspondence of the local LSBs appears to
be the damped $L_{\alpha}$ system (Wolfe 2001), which also
possesses disk-like morphology with large scale length, and a
constituting mass component which is gas rich and metal poor.

As the proto galaxy disk contracts and condenses, the global
spiral instability develops, which begins the mass-redistribution
process that transforms the flat surface density distribution to
a more and more centrally concentrated mass distribution.
The intermediate stage of this evolution resembles Freeman's
type II disks (Freeman 1970), and the later stage with a a 1/R projected
surface density resembles a Freeman's type I disk.

As the evolution progresses, the inward-accreted
disk stars rise out of the disk plane to become bulge stars. 
The thermalized three-dimensional bulge mass will 
have nearly constant volume density and undergoes solid-body rotation.  
The outer part of the visible disk starts to be drained of matter, 
producing the falling rotation curve.

This entire evolution sequence therefore corresponds to a
galaxy evoling from the lower rotation curve to the upper
rotation curve in Figure 1.

The twisted isophotes often found for the Hubble Deep Field
galaxies are expected to evolve into the observed variation of 
major axis with radius for elliptical galaxies
(Faber \& Jackson 1976), possibly bypassing the disk formation
phase, at least for some galaxies.  Like the spiral structure, 
the twisted isophotes are results of dissipation and differential rotation,
and just as in the case of spiral structure, these three-dimensional 
skewed structures will also lead to a phaseshift between
the potential and density distributions, resulting in accelerated
secular morphological changes. 

\section{Generation and Evolution of the Galaxy Scaling Relations
Along the Hubble Sequence}

Galaxy scaling relations (Faber \& Jackson 1976; Tully \& Fisher 1977;
Djorgovski \& Davis 1987; Dressler et al. 1987)
own their existence first and foremost
to the fact that galaxies are equilibrium configurations
and satisfy the Virial theorem relation (see, e.g., Binney
\& Tremaine 1987).  This relation dictates that during any quasi-equilibrium
evolution process of a gravitational system, $-E \sim T \sim -V/2$ where T
is the kinetic energy, V is the potential energy, and $E=T+V$ is the
total energy of the system.  The fact that spiral galaxies can remain
on the Tully-Fisher relation during secular morphological
evolution despite the fact that stars can't radiate about 1/2
of the converted potential energy away, is due to the
fact that spiral density wave actually carries about 1/2 of the
dissipated orbital energy (equal to potential energy) away to
the outer disk to be deposited there (Zhang 1999).  The existence of a
collective dissipation process at the spiral arms also made possible
the regulated conversion of orbital velocity to random velocity,
so a single kinetic energy (the half that is not being carried
away) can enter the Virial relation
no matter how it is apportioned between the circular
and the random motions.

According to the Virial theorem (we set G=1 in
the following derivations of scaling relations)
$
M_{dyn} = V_e^2 R_e,
$
where $M_{dyn}$ is the dynamical mass of a galaxy, 
$V_e$ is the effective velocity spread
which corresponds to the total velocity dispersion
for an elliptical galaxy or the maximum rotation velocity
for a spiral galaxy,
and $R_e$ is an effective radius which makes the Virial relation hold.

Define the mean surface brightness of a galaxy $SB$ as
$
SB \equiv {{L} \over {R_e^2}}
$
where $L$ is the luminosity of the galaxy, and the 
dynamical mass-to-light ratio $M_{dyn}/L$ can be written as
$
M_{dyn} \equiv (M_{dyn}/L) * L.
$
Therefore it follows that
\begin{equation}
(M_{dyn}/L)^2 * L^2 = M_{dyn}^2
= V_e^4 R^2 = V_e^4 {{L} \over {SB}},
\end{equation}
or
\begin{equation}
(M_{dyn}/L)^2 L = {{V_e^4} \over {SB}}
.
\end{equation}
Therefore we finally have from 
\begin{equation}
L = V_e^4 {{1} \over {SB}} {{1} \over {(M_{dyn}/L)^2}}.
\end{equation}
We thus see that in order to obtain the classical 
Tully-Fisher relation, which has $L \propto V_e^4$,
we must have $SB \cdot (M_{dyn}/L) \sim$ constant.  This
can be accomplished in two ways: to have $SB$ and
$M_{dyn}/L$ each being constant; or, to have the variations 
of the two factors offset each other 
during an evolution process.

Traditionally, the former was assumed to be the case (see, e.g.
the discussions and references in Shu 1982).  
It is now known that both these quantities vary considerably
for galaxies along the Hubble sequence.  The surface brightness
is found to be higher for earlier type galaxies (McGaugh \& de Blok 1998
and the references therein), whereas the dynamical mass-to-light is 
found to be lower for the earlier Hubble types  
(Zwaan et al. 1995; Bell \& de Jong 2001).
The observed opposing trends of variation of surface  brightness and
dynamical mass-to-light ratio is naturally explained as
the outcome of the secular baryonic mass accretion and the increase
fraction of luminous baryon mass, especially towards the central
region of a galaxy as its Huuble type evolve from late to early.
The Tully-Fisher relation can be maintained
as long as the increase in surface brightness is accompanied by 
a corresponding decrease in the dynamical mass to light ratio
during the secular evolution process. 

In the secular evolution scenario, since most of the elliptical 
galaxies (i.e. those lower luminosity, so-called disky ellipticals)
are the end results of
evolution from the initial condition of disks, we speculate
that spiral galaxies may satisfy similar kind of fundamental
plane relation just as ellipticals.  This has been found to be
so (see, e.g.,  Pharasyn, Simien, 
\& Heraudeau 1997).  The fundamental plane relation for spirals
can likewise be derived from the Virial theorem. Starting from 
\begin{equation}
V_e^2 = G {{M_{dyn}} \over R_e},
\end{equation}
we have 
\begin{equation}
V_e^2 = G {{M_{dyn}} \over L} { L \over R_e}
= L^{1/2} ({L \over R_e^2})^{1/2} {{M_{dyn}} \over L},
\end{equation}
which leads to
\begin{equation}
10 \log V_e = - (1+2\beta) M_t - \mu_e + constant
\label{eq:fp0}
,
\end{equation}
where $M_t$ is the absolute magnitude
and $\mu_e$ is the face-on average surface brightness in mag/arcsec$^2$,
and where we have assumed M$_{dyn}$/L $\propto L^{\beta}$.
Pharasyn et al. (1997) found that fitting I band and K band data
of a sample of spiral galaxies to this relation resulted in
$\beta \approx -0.15$, i.e. the mass-to-light ratio slightly
decreases with increasing luminosity, which is consistent
with what we would expect from a secular evolution picture,
i.e. as the secular evolution advances, L increases and M$_{dyn}$/L
decreases.  

The fundamental plane relation for elliptical galaxies
and bulges were first obtained by Faber et al. (1987), 
Djorgovski \& Davis (1987), and Dressler et al. (1987).
One type of the fundamental plane relation for
ellipticals can be written as 
(c.f. equation 2a and 2b of Djorgovski \& Davis 1987) 
\begin{equation}
L \sim V_e^{3.45} SB^{-0.86}.
\label{eq:fp2}
\end{equation}
This relation can be re-written into the form
\begin{equation}
M_t(R_e) = -8.62 (\log V_e + 0.1 \mu_e) + constant
,
\label{eq:fp3}
\end{equation}
which can be further written as
we arrive at
\begin{equation}
10 \log V_e = -1.25 M(R_e) - \mu_e + constant
.
\label{eq:fp4}
\end{equation}

Comparing equations (\ref{eq:fp0}) and (\ref{eq:fp4}), 
we see that apart from a possible
difference in the ``constant'' term (accounting for the
change in total luminosity), the only difference
between the spiral and elliptical fundamental plane
relations is in the different sign of the mass-to-light
ratio exponent $\beta$: $\beta=-0.15$ for spirals
and $\beta = 0.13$ for ellipticals and bulges, where
$\beta$ is defined through $M_{dyn}/L = L^{\beta}$.
This difference is apparently brought about by the fact
that spiral galaxies still have varying reserves of
baryonic dark matter to form stars, therefore as the
secular evolution proceeds (and therefore L increases)
the mass-to-light ratio decrease; whereas elliptical
galaxies have essentially exhausted their central
baryonic dark matter supply, thus the ellipticals in
more advanced stage of evolution (which also generally
have larger L) will experience greater degree of dimming,
which is reflected in the increase of mass-to-light
ratio with L.  

\smallskip

This research was supported in part
by funding from the Office of Naval Research.

\end{document}